\documentclass{aa}
\usepackage{txfonts} 

\usepackage{txfonts}
\usepackage{graphicx}

\def\p{\partial}
\def\a{\alpha}
\def\b{\beta}
\def\d{\delta}

\def\l{\kappa}
\def\r{\rho}
\def\g{\gamma}
\def\o{\omega}

\def\s{\sigma}

\def\O{\Omega}

\def\M{{\cal M}}
\def\R{{\cal R}}
\def\ra{\rightarrow}
\def\Ra{\Rightarrow}
\def\Lra{\Leftrightarrow}

\def\bm#1{\mbox{\boldmath{$#1$}}}

\begin{document}

\title{Perturbations of self-similar Bondi accretion}

\author{Jos\'e Gaite} 
\institute{Instituto de Matem{\'a}ticas y F{\'\i}sica Fundamental,
CSIC, Serrano 113bis, 28006 Madrid, Spain}

\date{Received: 24 August 2005 / Accepted: 12 November 2005}

\abstract{The question of stability of steady spherical accretion has
been studied for many years and, recently, the concept of {\em spatial
instability} has been introduced.
Here we study perturbations of steady spherical accretion flows (Bondi
solutions), restricting ourselves to the case of {\em self-similar}
flow, as a case amenable to analytic treatment and with
physical interest. We further restrict ourselves to its acoustic
perturbations.
The radial perturbation equation can be solved in terms of Bessel
functions.  We study the formulation of adequate boundary conditions
and decide for no matter-flux-perturbation conditions (at the Bondi
radius and at $r=0$).  We also consider the problem of initial
conditions and time evolution, in particular, of radial perturbations.
No spatial instability at $r=0$ is found. The time evolution is such
that perturbations eventually become {\em ergodic-like} and show no
trace of instability or of acquiring any remarkable pattern.
}

\maketitle


\section{Introduction}

The Bondi solutions for stationary spherical accretion onto a compact
object (Bondi, 1952) have been the basis for many analytical (or
semi-analytical) studies of spherical accretion.  Since there is a set
of solutions, it has been a moot point how to select the most adequate
among them. Bondi suggested that stability criteria would be useful
for this purpose. However, neither analytical studies (Garlick, 1979;
Petterson, Silk \& Ostriker, 1980) nor numerical studies (Ruffert,
1994) have detected any instatibility, in the sense that linear modes
have amplitudes increasing with time. Garlick's argument is very
general and powerful because it is based on energy conservation.
Kovalenko \& Eremin (1998) pointed out another possible type of
instability, namely, in space rather than in time. They stated the
nature of this instability as given by a perturbation whose amplitude
relative to the unperturbed quantity grows infinitely with decreasing
radius, so that one expects that it reaches the nonlinear regime.
Apart from this spatial notion of stability, Garlick (1979) had
proposed before ``to take a more sophisticated approach to the
stability problem'' (in the time domain): the perturbation,
independently of its global behaviour, could be increasing in a small
region (anywhere along the flow). However, he ruled this out as
``unphysical''. Garlick (1979) also considered briefly the initial
value problem.

Our purpose in this paper is to pay special attention to the problem
of time evolution of perturbations of Bondi accretion and to go as far
as we can with analytic methods. Given that the general problem of
initial conditions and time evolution is intrinsically difficult, we
are compelled to make a number of simplifications. First of all, we
will consider a particularly simple form of Bondi accretion:
self-similar Bondi accretion. It occurs for the value of the
polytropic index $\g = 5/3$, a limit case with physical interest (or
it holds approximately as an {\em intermediate asymptotic} limit for
$\g < 5/3$).  As an additional simplification, we will restrict
ourselves to acoustic perturbations, neglecting entropic perturbations
and vorticity. This is consistent, as generation of entropy and
vorticity are related, and they are simply advected with the flow 
and eventually vanish (Garlick, 1979). Moreover, to carry out a
complete study of time evolution we will restrict ourselves to radial
perturbations. In this case, the analytic treatment is simpler, as is
the graphical representation of the results.

We remark that a possible r\^ole of entropy and vorticity
perturbations has been explored recently (Foglizzo, 2001). However, in
accord with the traditional view, no instabilities can arise in
polytropic flow in absence of an external source of entropy or
vorticity. We consider no external sources.
 
We divide the paper in three main sections: The first one (numbered
second section) is a review of Bondi accretion and, furthermore, it
introduces the self-similar solutions. The next section is devoted to
the study of perturbations, in particular, acoustic modes, and to
formulate the appropriate boundary problem of differential equations.
We follow previous work but we try to be more systematic.
Unfortunately, the boundary problem does not lead to a Sturm-Liouville
problem, even though is connected with one.  The last main section
focuses on radial perturbations, like the papers of Petterson, Silk \&
Ostriker (1980) and of Theuns \& David (1992). In that section, we
precisely formulate and solve the corresponding eigenvalue problem of
ordinary differential equations (ODE's), finding an orthogonal system
of eigenfunctions. Hence we solve the problem of initial
conditions. We end by discussing the relevance of our results for the
problem of stability of Bondi accretion.  An appendix exposes the
asymptotic form of small wave-length perturbations (WKB
approximation).


\section{Steady spherical accretion: the Bondi solutions}
\label{sec:review}

The problem of steady spherical accretion is suitable for analytical
treatment and is the basis for more complicated and realistic problems
(Frank, King \& Raine, 1992; Shore, 1992).  
A compact object (e.g., a star) of mass
$M$ is at rest at the origin ($r=0$).  An infinite cloud of gas, which
is at rest at infinity and with density $\rho_\infty$ and pressure
$p_\infty$, is accreted by the compact object.  The motion of the gas
is assumed to be spherically symmetric and steady, so the fluid
partial differential equations (PDE's) become ODE's in the radius $r$.
Furthermore, the change of mass of the compact object, the effects of
radiation and viscosity are neglected.  Then the equations describing
the gas are
\begin{eqnarray}
\frac{d}{dr}(r^2 \rho {v}) = 0
\; ,
\label{density5-b}
\\
v \frac{d v}{dr} 
= - \frac{1}{\rho}\frac{d p}{dr}
- \frac{GM}{r^2}
\; .
\label{moment5-b}
\end{eqnarray}
with $\rho$ the gas density field, $v$ the gas velocity field, $p$ the
gas pressure and $G$ Newton's gravitational constant.  The continuity
equation (\ref{density5-b}) can be integrated immediately. We must add
an equation of state of the fluid, which we choose polytropic,
following Bondi (1952). Then we have a set of three equations:
\begin{eqnarray}
4 \pi r^2 \rho v&=& {\cal K}_1
\; ,
\label{density5-c}
\\
v \frac{d v}{d r}&=& -\frac{GM}{r^2}- \frac{1}{\rho} \frac{d p}{d r}
\; ,
\label{moment5-c}
\\
p &=&{\cal K}_2\, \rho^{\gamma}
\; .
\label{poly5-c}
\end{eqnarray}
with ${\cal K}_1$ and ${\cal K}_2$ constants (characterizing the
flow), and the polytropic exponent $1 \leq \gamma \leq 5/3$.  With a
suitable choice of $\g$, the polytropic equation is equivalent to
assuming that there is no heat transfer (adiabatic flow).  The speed
of sound, to be used later, is given by $c^2=d p/d\rho = {\cal K}_2
\g\,\rho^{\gamma-1}$.

Integrating the Euler equation (\ref{moment5-c}) over $r$ one gets
Bernouilli's equation.  Furthermore, the polytropic equation
(\ref{poly5-c}) allows one to integrate $dp$, so the problem boils
down to two algebraic equations, which can be expressed in
non-dimensional form by using the sound velocity in the gas at
infinity $c_\infty$. Let us define the non-dimensional variables
\begin{eqnarray}
x&=&\frac{r}{\R}
\; ,
\label{x}\\
y&=&\frac{\rho}{\rho_\infty}
\; ,
\\
u&=&\frac{v}{c_\infty}
\; ,
\\
\lambda&=& \frac{{\cal K}_1}{4 \pi \R^2 \rho_\infty c_\infty}
\;;
\end{eqnarray}
that is, $r$ is measured in units of the accretion (or Bondi) radius,
$\R = GM/c^2_\infty$, $v$ in units of the speed of sound $c_\infty$,
and $\r$ in units of $\r_\infty$.  The non-dimensional form of the
equations of motion is:
\begin{eqnarray}
x^2 y u &=& \lambda
\, ,
\\
\frac{u^2}{2} + \frac{y^{\g -1}-1}{\g-1} &=& \frac{1}{x}
\, .
\end{eqnarray}
Solving for $u$ in the first equation and substituting in the second,
we obtain
\begin{equation}
\frac{\lambda^2}{2}x^{-4} y^{-2} + n(y^{1/n}-1)= x^{-1}
\; ,
\label{denseq}
\end{equation}
where we have introduced the adiabatic index $n = 1/(\g-1)$.  As an
equation for $x$, it is a quartic algebraic equation, so it admits an
algebraic expression for its solutions (however complex, see Theuns \&
David, 1992).

\subsection{Self-similar Bondi solutions}
\label{sec:bondi}

The preceding algebraic relation between $y$ and $x$ (\ref{denseq})
can be simplified in some cases.  Let us assume first that the radial
distance is small, namely, that $x^{-1} \gg 1$, so that the term $n$
in Eq.~(\ref{denseq}) can be neglected (in comparison with $x^{-1}$).
The condition that the density is large, {\it i.e.,} $y \gg 1$, is
equivalent (assuming that $\g$ is not too close to one). Then we have
\begin{equation}
\frac{\lambda^2}{2}x^{-3} y^{-2} + n\,y^{1/n} x= 1
\; .
\label{denseq-s}
\end{equation}
For convenience, we define the variable $z = y^{1/n} x$ and rewrite 
this equation as
\begin{equation}
\frac{\lambda^2}{2}x^{-3+2n} z^{-2n} + n\, z= 1
\; .
\label{denseq-s2}
\end{equation}
This is still a complicated relation between $x$ and $z$, so we 
must look for further simplification. 

We see that in the particular case $n=3/2$, Eq.\ (\ref{denseq-s2}) is
just an algebraic equation in the sole variable $z$, with a solution
that only depends on the non-dimensional accretion rate
$\lambda$. Therefore, in this case, the non-dimensional density takes
the form $y = z(\lambda)^{3/2}\, x^{-3/2}$ and the non-dimensional
velocity takes the form $u = \lambda \,z(\lambda)^{-3/2}\,
x^{-1/2}$. We see that both density and velocity are {\em power laws}
of the radius.  More complicated solutions of Eq.~(\ref{denseq-s2})
can be considered but we will focus on the case $n=3/2 \Ra \g=5/3$,
which provides simple solutions and is realistic since it corresponds
to adiabatic flow of perfect monoatomic gases.

Then we write
\begin{eqnarray}
\rho(r) &=& \alpha\, r^{-3/2}
\; ,
\label{eq:rho-bondi}
\\
v(r) &=& \beta \,r^{-1/2}
\; ,
\label{eq:v-bondi}
\end{eqnarray}
with $\alpha$ and $\beta$ such that 
\begin{eqnarray}
\alpha \beta &=&\frac{{\cal K}_1}{4 \pi}
\; .
\end{eqnarray}
In this case, the speed of sound is 
\begin{eqnarray}
c^2 (r)= {\cal K}_2\, \frac{5}{3}\,\alpha^{2/3} r^{-1}= \sigma^2 r^{-1}
\,,
\end{eqnarray}
so the Mach number is given by
\begin{equation}
{\cal M} = \frac{v(r)}{c(r)}= \frac{\beta}{\sigma}
\, ,
\end{equation}
which is constant. So each solution is charaterized by either 
$\{\alpha, \beta, \s\}$ or $\{{\cal K}_1,{\cal K}_2, {\cal M}\}.$

We remark that self-similar Bondi solutions have no discontinuities
and are either subsonic or supersonic everywhere. On account of the
boundary condition at infinity, it is natural to discard the
supersonic self-similar solutions.

\subsubsection{Intermediate asymptotic limit when $\g < 5/3$}

It is commonly observed that nonlinear equations that do not have
similarity solutions at the outset develop them in an {\em
intermediate asymptotic} regime, that is, a regime between two very
different scales where both scales can be neglected (Barenblatt,
1996). In the problem of steady spherical accretion by a compact
object, we have two fundamental scales, namely, the Bondi radius $\R$
and the sonic radius. The latter is defined by the implicit equation
\begin{equation}
{r_s} = \frac{GM}{2 c^2(r_s)}
\end{equation}
(Frank, King \& Raine, 1992; Shore, 1992). This equation can be
written in non-dimensional variables as
\begin{equation}
{z_s(x_s)} = y_s(x_s)^{1/n} x_s = \frac{1}{2}\;,
\label{zs}
\end{equation}
where $y_s(x_s)$ is the solution of Eq.\ (\ref{denseq}) 
corresponding to $x_s$. Writing Eq.\ (\ref{denseq}) 
in terms of $x_s$ and $z_s = 1/2$, we have
\begin{equation}
{2^{2n-1}}\lambda^2\,x_s^{-3+2n} = n\,x_s + 1 -\frac{n}{2} 
\; .
\end{equation}
If $2 > n > 3/2 \Lra 3/2 < \g < 5/3$ this equation has zero, one or
two solutions, dependig on the value of the non-dimensional accretion
rate $\lambda$ (given $n$).  As $\lambda$ increases, the number of
solutions increases as well, being one at the critical value
$\lambda_\textrm{max}$, which is only function of $n$. This critical
value is called $\lambda_\textrm{max}$ because it is the maximal value
such that the flow is continuous (Theuns \& David, 1992).  The value
$$x_s(\lambda_\textrm{max}) = \frac{1}{2} - \frac{3}{4n}$$ is only
function of $n$ and approaches zero as $n \ra 3/2$.

If $n > 3/2$ but not much larger ($\g < 5/3$ but not much smaller),
and $\lambda$ is not much smaller than $\lambda_\textrm{max}$, the
solution of Eq.\ (\ref{denseq-s2}) for $z$ will have a dependence on
$x$ but only logarithmic, such that the power laws
(\ref{eq:rho-bondi}) and (\ref{eq:v-bondi}) will hold with corrections
proportional to $\log (r/r_s)$.  In other words, in the intermediate
asymptotics $r_s \ll r \ll \R$, the power laws are approximately
correct (so the Mach number ${\cal M}$ hardly grows with the flow).

For definiteness, we will set $\g = 5/3$ in the following, keeping in
mind that the results are of somewhat wider applicability.


\section{Linear perturbations of the self-similar Bondi solutions}
\label{sec:lin-pert}

Let us assume that the steady spherically symmetric solutions are
perturbed, such that the total velocity and density fields become
$\rho(r)+ \delta \rho({\bm x},t)$ and ${\bm v}(r) + \delta {\bm
v}({\bm x},t)$, with $\rho(r)$ and ${\bm v}(r)=v(r) \hat {\bm r}$ the
self-similar solutions, given by Eqs.~(\ref{eq:rho-bondi})
and~(\ref{eq:v-bondi}) in the self-similar case ($\hat{\bm r}$ is the
unit radial vector).

The independent perturbations are $\d \r$, $\d {\bm v}$ and the
``entropy perturbation'' $\d s$, where the ``entropy'' is defined as
$$s = \log \frac{c^2}{\g\,\r^{\g-1}}$$ (Kovalenko and Eremin, 1998).
We shall assume that the perturbations are such that $\d s = 0$, so
that they keep the polytropic equation of state. In particular, they
may be adiabatic perturbations; we shall henceforth denote
thermodynamic functions as if this is the case.  Then the equations
for linear perturbations are
\begin{eqnarray}
\frac{\partial \delta \rho}{\partial t} + \nabla \cdot
(\delta \rho \; {\bm v} + \rho  \; {\delta {\bm v}}) &=& 0
\; ,
\label{lin-non-1a}
\\
\frac{\partial {\delta {\bm v}}}{\partial t} 
+
({\bm v}\cdot\nabla)\; \delta {\bm v}
+
(\delta {\bm v}\cdot\nabla)\; {\bm v}
&=& - \nabla \delta \chi
\; ,
\label{lin-non-1b}
\end{eqnarray}
where 
\begin{equation}
\delta \chi = \frac{c^2 \,\delta \rho}{\rho}
\label{enthal}
\end{equation}
is the enthalpy perturbation (the speed of sound is given by $c^2
= {\cal K}_2 \g \r^{\g-1}$). Therefore, we have a system of four
linear PDE's.


\subsection{Solution for acoustic modes}
\label{sec:acoustic}

For the acoustic modes, $\d s = 0$ and the perturbation in the
velocity field can be written as $\delta {\bm v}= \nabla \phi.$ In
this case, and given that Bondi flows are curl free, we can write
Eq.~(\ref{lin-non-1b}) as the following differential equation for the
scalar potential field $\phi$
\begin{eqnarray}
\nabla \frac{\partial {\phi}}{\partial t} + \nabla ({\bm v} \cdot
\nabla \phi) &=& - \nabla \delta \chi \; .
\label{acoustic1}
\end{eqnarray}
This equation can be integrated to obtain (Kovalenko \& Eremin, 1998)
\begin{eqnarray}
\frac{\partial {\phi}}{\partial t} + {\bm v} \cdot \nabla \phi &=& -
\delta \chi \; ,
\label{acoustic2}
\end{eqnarray}
which provides $\delta \chi$ and hence $\delta \rho$ in terms of
$\phi$. Equation~(\ref{lin-non-1a}) can then be written in terms of
$\phi$ as follows
\begin{eqnarray}
\left[
\frac{\partial}{\partial t} +  {\bm v} \cdot \nabla +
\frac{c^2}{\rho} \nabla \cdot \left( \frac{\rho {\bm v}}{c^2}
\right)\right] \left( \frac{\partial}{\partial t} + {\bm v} \cdot \nabla 
\right) \phi
= \nonumber\\ 
\frac{c^2}{\rho} \nabla \cdot (\rho \nabla \phi)
\; 
\label{eq:diff-eq-phi-1}
\end{eqnarray}
[Eq.~(24) of Kovalenko \& Eremin (1998)]. This is an equation for
 non-homogeneous wave propagation.

One tries a solution of equation~(\ref{eq:diff-eq-phi-1}) for the
scalar potential $\phi$ by separation of the variables in spherical
coordinates, that is, a solution in the (complex) form
\begin{eqnarray}
\phi(r,\theta,\varphi,t)= R(r)\, Y_{lm}(\theta,\varphi)\,e^{-i\o t}
\, .
\label{eq:phi-sol-1}
\end{eqnarray}
Given this form of the scalar potential, the velocity perturbation is
given by
\begin{eqnarray*}
\d{\bm v} = (\d v_r, \d v_\theta, \d v_\varphi)
\, ,
\end{eqnarray*}
with
\begin{eqnarray}
\d v_r (r,\theta,\varphi,t)
&=& \textrm{Re}[e^{-i\o t} \frac{dR(r)}{dr} Y_{lm}(\theta,\varphi)]
\; ,
\\
\d v_\theta  (r,\theta,\varphi,t)&=&
\textrm{Re}[e^{-i\o t}\frac{R(r)}{r} \partial_\theta Y_{lm}(\theta,\varphi)]
\; ,
\\
\d v_\varphi (r,\theta,\varphi,t)
&=&\textrm{Re}[e^{-i\o t}\frac{R(r)}{r \sin \theta} \partial_\varphi Y_{lm}
(\theta,\varphi)]
\; 
\end{eqnarray}
(after taking the real part of the complex quantities).  Note that the
angular components form a {\em poloidal} vector field and the form of
the radial component ensures the absence of vorticity (Garlick, 1979).

The differential equation satisfied by $R(r)$ is 
\begin{eqnarray}
(\sigma^2-\beta^2)r^2 R'' + (\frac{\sigma^2-\beta^2}{2} + 2 i \beta \o
r^{3/2}) r R' + \nonumber\\ \left[-l(l+1)\sigma^2 + \o^2 r^{3} + i
\beta \o r^{3/2}\right] R = 0 \, .
\label{eq:radial-diff-1}
\end{eqnarray}
Following standard methods of the theory of ODE's, the second order
equation
\begin{eqnarray}
R''(r)+ P(r) R'(r) +Q(r) R(r) = 0\, ,
\label{diff-1}
\end{eqnarray}
with
\begin{eqnarray}
P(r)&=&\frac{1}{2r} +\frac{2i\b\o r^{1/2}}{\sigma^2-\beta^2} \\
Q(r)&=&\frac{\o^2 r - l(l+1) \s^2 r^{-2} + i\b \o
r^{-1/2}}{\sigma^2-\beta^2} \; .
\end{eqnarray}
can be simplified by the following change of variable:
\begin{eqnarray}
\log R(r) = \log h(r) - \frac{1}{2}\int dr P(r)\, ,
\end{eqnarray}
which integrated yields 
\begin{eqnarray}
R(r) = r^{-1/4} \exp{\frac{-2i\beta \o r^{3/2}}{3(\sigma^2-\beta^2)}
}\,  h(r)\,.
\end{eqnarray}
In terms of this new dependent variable $h(r)$, the differential
equation (\ref{eq:radial-diff-1}) becomes
\begin{eqnarray}
h''(r) + (a_1 r^{-2} + a_2 r) h(r) = 0 \; ,
\label{selfadj-diffeq}
\end{eqnarray}
with
\begin{eqnarray}
a_1&=&-\frac{3}{16} - \frac{l(l+1)\s^2}{\sigma^2-\beta^2}\,,
\\
a_2&=&\left(\frac{\s\o}{\sigma^2-\beta^2}\right)^2\,.
\end{eqnarray}

The preceding differential equation (\ref{selfadj-diffeq}) has one
regular singular point at $r=0$ and one irregular singular point at
$r=\infty$. In fact, it can be transformed into the Bessel equation by
making $r^{3/2}$ the dependent variable.  So we obtain
\begin{eqnarray}
R(r) = r^{1/4} e^{-i \mu r^{3/2}} [C_1 J_\nu (\l r^{3/2}) 
+ C_2 N_\nu(\l r^{3/2})]
\; ,
\label{solutionR}
\end{eqnarray}
where
\begin{eqnarray*} 
\mu =\frac{2 \beta \o}{3(\sigma^2-\beta^2)},\\ 
\l= \frac{2\sigma\o}{3 (\sigma^2 - \beta^2)},\\ \nu
=\frac{\sqrt{[1+16l(l+1)]\sigma^2
-\beta^2}}{6\sqrt{\sigma^2-\beta^2}},
\end{eqnarray*}
and $C_1,C_2$ are, in general, two arbitrary complex numbers.

Note that the order $\nu$ of the Bessel function in solution
(\ref{solutionR}) does not depend on $\b$ and $\s$ separately but on
its ratio, that is, the Mach number, ${\cal M} = \beta/\sigma$ (in
addition to depending on the angular number $l$).  If the values of
${\cal M}$ and $l$ are generic such that $\nu$ is non-integer, we can
write the solution of the equation as
\begin{eqnarray}
R(r) = r^{1/4} e^{-i \mu r^{3/2}} [C_1 J_\nu (\l r^{3/2}) 
+ C_2 J_{-\nu}(\l r^{3/2})]
\; .
\label{solution2R}
\end{eqnarray}
In this form, the most general (complex) 
solution of the scalar potential is
\begin{eqnarray}
\phi(r,\theta,\varphi,t)= 
e^{-i\o t} \; Y_{lm} (\theta, \varphi)\,
r^{1/4} e^{-i \mu r^{3/2}} 
\times \nonumber\\
\left(C_1 J_\nu (\l r^{3/2}) + C_2 J_{-\nu}(\l r^{3/2})\right).
\label{potential}
\end{eqnarray}
Since we assume $\M < 1$, the signs of $\mu$ and $\l$ are positive
and $\nu$ is real.

Boundary conditions are needed in order to determine both $C_1$ and
$C_2$ and the $\o$-spectrum.  Notice that the constants $C_1$ and
$C_2$ need to be specified for each solution, that is, they are
indexed by $l,m$ and the $\o$-spectrum index.

\subsection{Boundary conditions}
\label{bc-sec}

The acoustic perturbations satisfy the wave equation
(\ref{eq:diff-eq-phi-1}), so a suitable boundary condition on a
surface is needed.  After the separation of variables in spherical
coordinates, one needs to impose two boundary conditions on the radial
differential equation, as this equation is of second order (Morse \&
Feshbach, 1953).  So one assumes that the boundary conditions adapt to
the spherical coordinates and, in particular, the boundaries are two
spherical surfaces of inner radius $r_1$ and outer radius $r_2$,
respectively.  The natural inner radius is either the accretor or the
sonic radius.  The natural candidate for outer radius is the Bondi (or
accretion) radius ${\cal R}$: as $r$ gets larger than ${\cal R}$, the
thermal energy of the nearly homogeneous gas becomes dominant over its
gravitational energy and we assume that this homogeneous gas remains
unperturbed.

The choice of boundary conditions of Petterson, Silk \& Ostriker
(1980) (for radial perturbations) is that the flux perturbation
$\delta f$ should vanish at both $r_1$ and $r_2$: since the flux of
matter through a sphere of radius $r$ is constant in the Bondi
solutions [Eq.\ (\ref{density5-c})], it seems natural to impose that
the perturbations do not change it on the inner and outer boundary.
Unlike for radial modes, for general acoustic modes the variation of
the matter flux current $\bm{j} = \rho{\bm v}$, namely, $\delta \bm{j}
= \rho \,\delta {\bm v} + \delta \rho \,{\bm v}\,,$ depends on the
angles $\theta,\varphi$.  The perturbed flux of matter through a
sphere of radius $r$ is given by the angular integral of its radial
component, proportional to $\int d\O\,Y_{lm}(\theta,\varphi)$.  This
angular integral vanishes if $l \neq 0$, so this condition is only
useful for radial modes.

Instead, Kovalenko \& Eremin (1998) derive their boundary conditions
from the law of conservation of acoustic energy
\begin{equation}
\frac{\p E}{\p t} = \nabla\cdot \bm{W} \,,
\end{equation}
where the acoustic energy density is
\begin{equation}
E = \frac{\rho}{2c^2} 
\left[\left(\frac{\p \phi}{\p t}\right)^2 +
{c^2}(\nabla\phi)^2  - (\bm{v}\cdot\nabla \phi)^2 \right],
\label{E}
\end{equation}
and the acoustic energy flux current is
\begin{equation}
\bm{W} = \frac{\rho}{c^2} \frac{\p \phi}{\p t} 
\left[\bm{v}\, \frac{\p \phi}{\p t} - {c^2} 
\nabla\phi  + \bm{v}\cdot(\bm{v}\cdot\nabla \phi) \right]
= - \p_t\phi \; \delta \bm{j}\,.
\label{W}
\end{equation}
The acoustic energy flux through a spherical surface centred on the 
origin is given by 
\begin{equation}
r^2 \int d\O\;W_r = 
\frac{r^2\rho}{c^2}\int d\O\; \p_t \phi \left[v\, \p_t \phi - {c^2} 
\p_r\phi  + v^2\p_r \phi \right].
\label{Wr}
\end{equation}
This quantity is quadratic in $\phi$ and does not necessarily vanish
for any $l$ upon performing the angular integral.

We see from Eq.~(\ref{Wr}) that sufficient conditions for the
vanishing of the acoustic energy flux are either
\begin{equation}
\p_t \phi = 0
\end{equation}
or
\begin{equation}
r^2\,\hat {\bm r} \cdot \delta \bm{j} = -\frac{r^2\rho}{c^2}
\left[v\, \p_t \phi - {c^2} \p_r\phi  + v^2\p_r \phi \right] = 0\,.
\label{bc}
\end{equation}
The latter condition implies, of course, the vanishing of its integral
$\delta f$ over any angular domain, so it is the appropriate
generalization of the condition of Petterson, Silk \& Ostriker (1980).
Hence, we choose it as our boundary condition; namely, this quantity
should vanish at both $r_1$ and $r_2$.  It can be expressed only in
terms of $R(r)$ as
\begin{eqnarray}
\frac{r^2\rho}{c^2}\left[i \o v R +  (c^2 - v^2) R'\right]  = 0
\; .
\label{bc-R}
\end{eqnarray}

The discussion above is general, that is, it is not restricted to
self-similar solutions. For self-similar solutions, the accretor and
sonic radii vanish, so we take $r_1 = 0$; on the other hand, the 
Bondi radius ${\cal R} \ra \infty$. However, we need a finite outer
radius to have a {\em discrete} spectrum, so we take $r_2 = \R$
finite, keeping in mind that the actual spectrum becomes
continuous in the limit ${\cal R} \ra \infty$.

The radial equation (\ref{eq:radial-diff-1}) for $R(r)$ is not
self-adjoint [the subject of adjointness of ODE's and the
Sturm-Liouville boundary problem is treated, for example, by Morse \&
Feshbach (1953) and Coddington \& Levinson (1955)].  However, the
transformed equation (\ref{selfadj-diffeq}) for $h(r)$ is patently
self-adjoint. The boundary condition (\ref{bc-R}) in terms of $h(r)$
reads
\begin{eqnarray}
r^{-3/4}[-h +  4 r h'(r)]  = 0\, 
\label{bc-h}
\end{eqnarray}
(at $r_1$ and $r_2$).  Therefore, the boundary problem for $h(r)$ is
of Sturm-Liouville type and the eigenvalues and eigenfunctions fulfill
the corresponding properties; in particular, the eigenfunctions are
orthogonal (Morse \& Feshbach, 1953). This is natural, because $h(r)$
is a combination of Bessel functions (with a factor $r^{1/2}$). In
contrast, the boundary problem for $R(r)$ is {\em not} of
Sturm-Liouville type and the eigenfunctions need not be orthogonal
(note that the $\l$ dependence of $\mu$ in Eq.\ (\ref{solutionR})
spoils the orthogonality inherent to the Bessel functions). Having
an orthogonal system of eigenfunctions is important for the
problem of initial conditions, namely, for finding the coefficients of
the expansion corresponding to the initial data.

We now apply Eq.\ (\ref{bc-h}) at $r_1=0$ and $r_2$. 
For the former, we need to take the limit $r \ra 0$ of 
Eq.\ (\ref{bc-h}). Writing it in terms of Bessel functions, 
with the dependent variable $z= \l r^{3/2}$, we have
\begin{eqnarray}
\lim_{z \ra 0} z^{-1/6}\left(C_1[J_\nu(z) + 6 z  J'_\nu(z) ] + 
\right.\nonumber\\
\left. C_2[J_{-\nu}(z) + 6 z  J'_{-\nu}(z) ]\right) &=& 0
\; .
\end{eqnarray}
In this limit, it is sufficient to consider the first term 
of the Taylor expansion of the Bessel functions, namely,
\begin{eqnarray}
z^{-1/6}
[J_\nu(z) + 6 z  J'_\nu(z)] \approx \frac{2^{-\nu}}{\Gamma(\nu + 1)}
(1+ 6 \nu) z^{\nu-1/6} 
\, .
\end{eqnarray}
When $\nu > 1/6$ ($l>0$) we are led to imposing $C_2 = 0$ 
to remove negative exponents (divergent as $r \ra 0$). If
$\nu = 1/6$ (corresponding to the radial modes), we must
take $C_1 = 0$ instead.

The outer boundary condition at $r_2$ yields
\begin{eqnarray}
J_\nu (z_2) + 6 z_2  J'_\nu (z_2) = 0
\label{boundary2b}
\, ,
\end{eqnarray}
which provides the allowed eigenvalues $\l_n$, and hence 
the $\o_n$.

Finally, it is important to analyse the relative variations of the
density and the velocity.  Their behaviour in the limit $r \ra 0$ is:
\begin{eqnarray}
\frac{\d v_r}{v} \sim 
r^{1/2} R'(r) \sim r^{-1/4 + 3\nu/2}\,, 
\label{dvr} \\
\frac{\d v_\theta}{v} \sim 
\frac{\d v_\varphi}{v} \sim 
r^{1/2} \frac{R(r)}{r} \sim r^{-1/4 + 3\nu/2}\,, 
\label{dvt-dvf} \\
\frac{\d \r}{\r} = \frac{1}{c^2}[i\o\,R(r) - v R'(r)]
\sim r^{-1/4 + 3\nu/2}
\,, 
\label{dr}
\end{eqnarray}
for $\nu > 1/6$; for $\nu = 1/6$, ${\d v_r}/{v} \sim {\d \r}/{\r} \sim
r$. All these relative variations vanish as $r \ra 0$, so there are
{\em no spatial instabilities} at the inner radius and linear
perturbation theory is sound.

\section{Radial perturbations}

The simplest acoustic modes are the radial perturbations, studied by
Petterson, Silk \& Ostriker (1980) and Theuns \& David (1992), for
general Bondi flow.  Indeed, a more thorough analytical treatment is
also possible in the self-similar case if we restrict ourselves to
radial perturbations.  Petterson, Silk \& Ostriker (1980) obtained
general stability results: the problem of standing waves with
vanishing flux boundary conditions has only real eigenvalues of $\o$,
and radially travelling waves can be shown not to grow (by an energy
argument). Theuns \& David (1992) derived an associated
Sturm-Liouville boundary problem by decomposing the flux perturbation
$\d f$ as if it were composed of modulated waves with a position
dependent phase velocity. This decomposition leads to a self-adjoint
equation for the amplitude.  Their procedure is analogous to the
transformation that led us from the radial Eq.\
(\ref{eq:radial-diff-1}) to the self-adjoint equation
(\ref{selfadj-diffeq}).

Although the associated Sturm-Liouville boundary problem 
yields the eigenfunctions of the original boundary problem, it does not 
provide us with an orthogonality relation for them, which 
may exist or not. Fortunately, a specific mathematical treatment 
of the first order equations for radial perturbations allows us 
to find a definite scalar product under which the eigenfunctions
are orthogonal. This is exposed in appendix \ref{appendix_A}. 
Next, we show the solution to the first order boundary problem, 
with the orthogonality relation, and then we proceed to 
solve the problem of initial conditions to study 
the time evolution.

\subsection{Solution of the boundary problem}

The general linear perturbation equations (\ref{lin-non-1a}) and
(\ref{lin-non-1b}) restricted to radial modes give
\begin{eqnarray}
-i\o\,\delta \rho + \frac{1}{r^2}
\partial_r \left[ r^2 (\delta \rho \; v + \rho \; \delta v)\right]
 &=& 0\, ,
\label{F-rad-1a}
\\
-i\o\,\delta v + 
\partial_r\left(\frac{c^2}{\r}\,\d \r + v\,{\delta v}\right)
&=& 0\, .
\label{F-rad-1b}
\end{eqnarray}
We denote $y = (\d f,
\d g)$, where $\d f = r^2(\delta \rho \, v + \rho \, \delta v)$
is the perturbation of the flow 
(but for the factor $4\pi$) and 
$\d g = (c^2/\r)\,\d \r + v\,{\delta v} = i\o \phi$ 
[according to Eq.\ (\ref{F-rad-1b})].
Solution (\ref{solution2R}) with $\nu = 1/6$ and $C_1 = 0$
(because of the boundary condition at $r=0$) gives
\begin{eqnarray}
\d f(r) =  \exp\left(-i\M\l r^{3/2}\right)
r^{5/4}\, i \, J_{5/6}(\l r^{3/2})\, ,
\label{f}
\\
\d g(r) =  \exp\left(-i\M\l r^{3/2}\right)
\frac{\sigma}{\a}\,
r^{1/4} \,J_{-1/6}(\l r^{3/2})\,,
\label{g}
\end{eqnarray}
where we have used the identity $J_{- 1/6}(z) + 6 z J'_{- 1/6}(z) = -
6 z J_{5/6}(z)$ to simplify $\d f$, and we have divided both
components by the constant total flux so that $\d f$ becomes
non-dimensional.  Now we impose the boundary condition $\d f = 0$ at
$r=1$ (we normalize the radial variable by dividing it by $r_2
=\R$). It determines that the posibles values of $\l$ are the zeros of
the Bessel function $J_{5/6}$ and so yields the eigenvalues $\l_n$.
Redefining the time unit such that $t \ra \s t$, we have the modes
\begin{eqnarray}
\d f_n(r,t) = \exp\left[-i\l_n\left(\M r^{3/2}+\frac{3}{2} (1
-\M^2)t\right)\right] \times \nonumber\\
r^{5/4}\, i \,
J_{5/6}(\l_n r^{3/2})\, ,
\label{fn}
\\ \d g_n(r,t) = \exp\left[-i\l_n\left(\M r^{3/2}+\frac{3}{2} (1
-\M^2)t\right)\right] \times \nonumber\\
\frac{\sigma}{\a}\,r^{1/4} \,J_{-1/6}(\l_n r^{3/2})\,,
\label{gn}
\end{eqnarray}
where $\{\l_n\}_{n=1}^\infty$ are the zeros of the Bessel function
$J_{5/6}$.  Furthermore, one can reverse the sign of $\l_n$, obtaining
the conjugate eigenfunction $y_n^* = (\d f_n^*, \d g_n^*)$, up to a
constant. Hence we denote $\l_{-n} = -\l_n$ and $y_{-n} = y_n^*$.  An
additional constant mode is necessary, namely, $y_0 = (0,1)$ (arising
from the limit of Eqs.~(\ref{f}) and (\ref{g}) as $\l \ra 0$).  It
corresponds to a constant velocity potential.  So the full set of
modes is $\{y_n\}_{n=-\infty}^\infty$.
 
These modes are orthogonal with respect to the product of
Eq.~(\ref{orthog}), that is,
\begin{eqnarray}
\langle y_n, y_m \rangle  = {\a}
\int_0^1 \frac{1}{\s^2 - \b^2} 
\left[\frac{\s^2}{\a} r^{-1/2} \d f_n^*\, \d f_m -
\right.\nonumber\\ \left.
 \b \,r^{1/2} (\d f_n^*\, \d g_m + \d g_n^*\, \d f_m) +
\a \,r^{3/2} \d g_n^* \,\d g_m \right]
dr \propto \delta_{nm}\,.
\label{orthog-J}
\end{eqnarray}
(To have a non-dimensional scalar product, we have introduced the
constant $\a$.)  We have checked numerically the orthogonality
relation (\ref{orthog-J}) for particular values of $\M$ and for modes
up to $n \simeq 100$.

\subsection{Initial conditions}

Having solved the boundary problem in terms of orthogonal modes, we
are in position to solve the problem of initial conditions.  As
initial conditions, we must take two functions $\d f(r,0)$ and $\d
g(r,0)$ that satisfy the boundary conditions, namely, $\d f(0,0) = \d
f(1,0) = 0$.  Then we expand $y = (\d f, \d g)$ in modes as
\begin{equation}
y(r) = \sum_{n=-\infty}^{\infty} c_n y_n(r)
\label{expan}
\end{equation}
Since 
$y_{-n} = y_n^*$, we require $c_{-n}
= c_n^*$ to have real $y$. So we can also write
$$
y(r) = c_0 y_0 + 2\, \textrm{Re}\left[\sum_{n=1}^{\infty} c_n
y_n(r)\right].
$$ Using the scalar product of Eq.~(\ref{orthog-J}), we can obtain the
coefficients in the expansion (\ref{expan}):
\begin{equation}
c_n = \frac{\langle y_n, y \rangle}{\langle y_n, y_n \rangle}
\,.
\label{coef}
\end{equation}
Either $(\d f(0,r),\d g(0,r))$ or the set of coefficients
$\{c_n\}_{n=0}^\infty$ contain the information on both density and
velocity perturbations and constitute alternate definitions of the
initial conditions.

\subsection{Time evolution}

Given the set of coefficients $\{c_n\}_{n=0}^\infty$, the solution of
the boundary problem in terms of the modes (\ref{fn}) and (\ref{gn})
can be written
\begin{equation}
y(r,t) = c_0 y_0 + 2\,\textrm{Re}\left[\sum_{n=1}^{\infty} 
c_n e^{-i\o_n t} y_n(r)\right].
\label{evol}
\end{equation}
Therefore, $y(t,r)$ adopts an expansion similar to the one 
of $y(0,r)$, Eq.\ (\ref{expan}), but with time-dependent coefficients
$$
c_n(t) = c_n e^{-i\o_n t}.
$$

Note that the norm
\begin{equation}
\langle y, y \rangle = \left|c_0\right|^2 \langle y_0, y_0 \rangle + 
2 \sum_{n=1}^{\infty} \left|c_n\right|^2 \langle y_n, y_n \rangle .
\label{power}
\end{equation}
is invariant in time. We can express this norm in terms of the
variables $\d\r$ and $\d v$ or, alternatively, in terms of $\p_t \phi
= -c^2 \d\r/\r - v \d v$ and $\d v$:
\begin{eqnarray}
\langle y, y \rangle = \int_0^1 \left(\frac{r^2 c^2}{\r} (\d\r)^2 +
2 r^2 v \d\r \d v + r^2 \r (\d v)^2 \right) dr \nonumber\\
= \int_0^1 \frac{r^2 \r}{c^2}  \left[ (\p_t \phi)^2 +
(c^2 - v^2) (\d v)^2 \right] dr\,.
\label{norm}
\end{eqnarray}
Comparison with the expression for the energy density (\ref{E}) shows
that the norm is just the energy (but for a constant factor), so it is
naturally invariant in time. Therefore, the mode decomposition of the
norm (\ref{power}) expresses the total energy as a weighted sum of the
energies of all modes.  The ``energy spectrum''
$\{\left|c_n\right|^2\}_{n=0}^\infty$ (invariant in time)
characterises the initial conditions in combination with the initial
phases of the coefficients.  Suitable initial conditions must be
normalizable, that is, must have finite energy.  From the asymptotic
form (\ref{orthog-J-as}), ${\langle y_n, y_n \rangle} \propto
1/(1+6n)$, we deduce that we must demand that $\lim_{n \ra \infty}
\left|c_n\right|^2 = 0$ (otherwise, the asymptotic form of the series
(\ref{power}) diverges like the harmonic series, at the least).

If we define the initial conditions by the energy spectrum and the
initial phases of the coefficients, the forms of $\d f$ and $\d g$
crucially depend on the correlation between those phases.  In general,
the evolution given by Eq.\ (\ref{evol}) is such that the correlation
between the phases decreases with time and is finally lost as $t \ra
\infty$. To be precise, this happens unless the eigenvalues $\o_n$
have a common ratio, in which case the solution is periodic. But this
does not happen in our case.  In other words, the type of motion that
we find is such that it leads to an evolution that is ergodic in a
restricted sense, that is, an evolution that conserves the energy
spectrum but which otherwise explores the available phase space.  A
typical state is given by taking the coefficient phases to be random.
It could happen that either $\d f(r,t)$ or $\d g(r,t)$ became
concentrated about a value of $r$ at some $t$ (and they could even
diverge) but it is extremely unlikely.

\begin{figure}
\centering{\includegraphics[width=8cm]{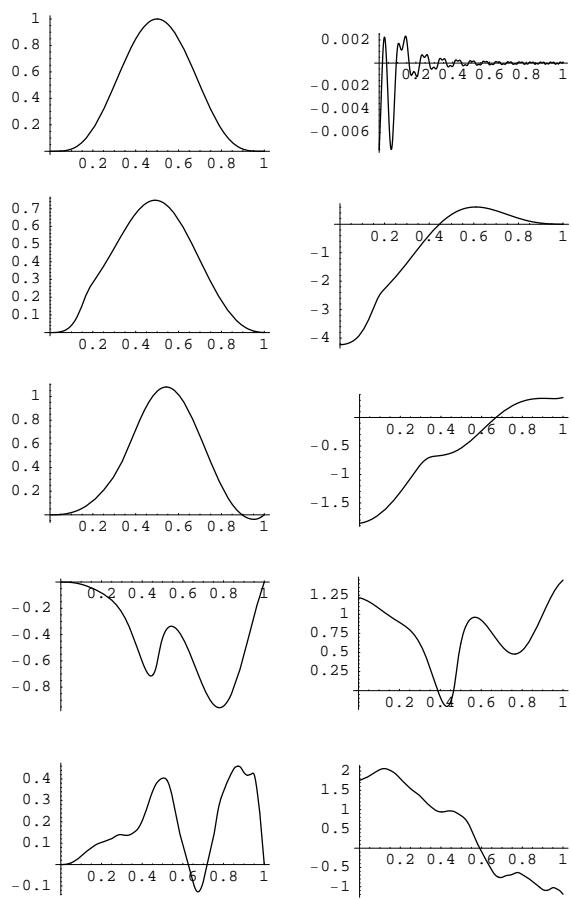}}
\caption{Time evolution of initial configuration $\d f(r,0) = b_4(r),
\d g(r,0) = 0$, with modes up to $n=50$ (corresponding to a relative
error in the norm $< 10^{-6}$). We have taken $\M=1/2$.  On the
left-hand side are plots of $\d f(r,t)$ and on the right-hand side are
plots of $\d g(r,t)$, for $t = 0, 0.1, 1.7, 100.$ The bottom plots
correspond to a realization of the same energy spectrum with random
phases.}
\label{t-evol}
\end{figure}

We have studied the evolution of particular radial perturbations of
self-similar Bondi flow with initial conditions that are sufficiently
smooth. They must have an energy spectrum that decays rapidly as $n
\ra \infty$. As this condition is time-invariant, smoothness is
preserved in time.  Functions such that they are concentrated in a
sort of bell shape have special interest. A
Gaussian profile is not allowed because of the boundary conditions
($\d f$ vanishes at $r=0,1$); but we can take polynomial bell-shaped
functions, for example, $b_n(x) = [4\,x(1-x)]^n, \:\:n = 1, 2, \dots$
(which are succesively more concentrated).  The time evolution of
initial configuration $\d f(r,0) = b_4(r), \d g(r,0) = 0$ is plotted
at several times in Fig.~\ref{t-evol}. This function gives rise to an
energy spectrum that decays very rapidly, so the modes with $n \leq
50$ correspond to a relative error in its norm (total energy) smaller
than $10^{-6}$.  We observe that $\d g$ initially evolves fast,
reaching a non-negligible value at $r=0$ in a very short time. We also
observe that $\d f(r,0)$ is almost reproduced at $t \simeq 1.7$ (but
$\d g(r,0) = 0$ does not seem to be reproduced). Nevertheless, there
is no periodicity and the reason why $\d f$ seems periodic is that the
power spectrum is very concentrated in a small number of modes that
are distributed with sufficient regularity.  At any rate, the phase
correlation among them is destroyed for longer times, as the following
plots in Fig.~\ref{t-evol} prove. The last couple of plots show a
``typical'' long-time state, with random phases.

\section{Discussion}

We have examined some of the most interesting questions regarding the
instability of Bondi accretion.  The first question is about the
spatial stability of perturbations, in the limit $r \ra 0$.  For our
self-similar solutions, the meaning of this limit is different from
the meaning in Kovalenko \& Eremin's (1998), because similarity
implies that the accretor and sonic radii vanish.  The behaviour of
perturbations in the limit $r \ra 0$ depends on the type of inner
boundary condition.  We have shown that the natural boundary condition
is the no-energy-flux perturbation condition, which is equivalent to
the vanishing of either the velocity potential time derivative or the
radial matter current variation.  We have focused on the latter
condition as the natural generalization of the boundary conditions of
Petterson, Silk \& Ostriker (1980) for radial perturbations.  Under
that condition, the relative variations of the physical variables,
velocity and density, are bounded in the limit $r \ra 0$.  Therefore,
the spatial stability of perturbations is ensured.  This conclusion
disagrees with an assertion by Kovalenko \& Eremin (1998), namely,
that some perturbations are spatially unstable, for example, a
non-radial subcritical mode. We attribute this difference to their
imposing the no-energy-flux condition only under time average (which
seems incomplete).

The second question is about the time evolution of perturbations.  We
have seen that the evolution is ergodic-like, namely, the
perturbations make a random walk in the section of the total phase
space with given energy spectrum, losing memory of the initial
conditions (the initial phases of the coefficients $c_n$).  Regarding
Garlick's proposal of perturbations that, independently of their
global behaviour, could be increasing in a small region, we can now
conclude that that can actually happen, but it is extremely unlikely:
indeed, we have shown that an initial condition concentrated in a
small region is forgotten in the evolution and not encountered again
in a random walk in the big phase space.  Furthermore, if we consider
the non-linear coupling among the modes, with the corresponding
transfer of energy, we deduce that the energy spectrum is not really
conserved, so the only truly conserved quantity is the total
energy. Therefore, in the long run, the energy spectrum thermalizes
and the evolution becomes fully ergodic. In addition, the presence of
viscosity (or other dissipative processes) would also transform the
energy of perturbations into thermal energy.  We see that an energy
concentration in a small region corresponds to a situation of entropy
spontaneously decreasing; so we fully agree with Garlick that this
possibility must be ruled out as unphysical.

We have restricted ourselves to acoustic perturbations of self-similar
Bondi accretion, but many results must be applicable to
non-self-similar Bondi accretion.  For example, the stability and
ergodic-like nature of the evolution of acoustic perturbations,
ultimately due to the reality of the spectrum of eigenvalues $\o_n$
and to its genericity (in the sense of not having a uniform
distribution).


\begin{acknowledgements}
I am grateful to Carmen Molina-Par{\'\i}s for conversations.  My work
is supported by the ``Ram\'on y Cajal'' program and by grant
BFM2002-01014 of the Ministerio de Edu\-caci\'on y Ciencia.
\end{acknowledgements}


\appendix

\section{Eigenvalue problem for radial perturbations}
\label{appendix_A}

The perturbation equations (\ref{lin-non-1a}) and
(\ref{lin-non-1b}) become for radial modes 
\begin{eqnarray}
\partial_t \delta \rho + \frac{1}{r^2}
\partial_r \left[ r^2 (\delta \rho \; v + \rho \; \delta v)\right]
 &=& 0\, ,
\label{rad-1a}
\\
\partial_t {\delta v} +
\partial_r\left(\frac{c^2}{\r}\,\d \r + v\,{\delta v}\right)
&=& 0\, .
\label{rad-1b}
\end{eqnarray}
After performing the Fourier transform for the time variable,
\begin{eqnarray}
-i\o\,\delta \rho + \frac{1}{r^2}
\partial_r \left[ r^2 (\delta \rho \; v + \rho \; \delta v)\right]
 &=& 0\, ,
\\
-i\o\,\delta v + 
\partial_r\left(\frac{c^2}{\r}\,\d \r + v\,{\delta v}\right)
&=& 0\, .
\end{eqnarray}
We define the vector $x = (r^2 \delta \rho, \delta v)$ and the matrix
$$
A = \left(
\begin{array}{cc}
v(r) & r^2\,\rho (r) \\ \frac{{c(r)}^2}{r^2\,\rho (r)} & v(r)    
\end{array}
\right),
$$ such that $\det A = v^2 - c^2 \neq 0.$ Then we can write the ODE's
(\ref{F-rad-1a}) and (\ref{F-rad-1b}) as the vector equation
\begin{equation}
\frac{d(A\cdot x)}{dr} = i\o\,x\,.  
\label{vec-eq}
\end{equation}
Its adjoint equation is 
\begin{equation}
A^\dag\cdot\frac{d {\bar y}}{dr} = i\o^*\,{\bar y}\,,  
\label{ad-vec-eq}
\end{equation}
where the adjoint matrix $A^\dag$ is, in this case, just the 
transpose of $A$, and we have written the unknown vector 
function as ${\bar y}$ for later convenience. The derivation of 
this adjoint equation holds as long as the boundary conditions 
are such that ${\bar y}^*\cdot A x$ vanishes at the ends of 
the interval of integration that defines the scalar product
of functions.

According to the general theory of ODE's (Coddington \& Levinson,
1955), the eigenfunctions of the original ODE and the eigenfunctions
of its adjoint equation are orthogonal, that is,
\begin{equation}
\int  {{\bar y}_n}^* \cdot x_m \,dr \propto \delta_{nm}\,,
\label{orthog0}
\end{equation}
and their respective eigenvalues are conjugate.  But the eigenvalues
are in fact real in our case due to a property of the matrix $A$,
namely, it is transformed into its adjoint by the interchange of the
two components of vectors.  To make this property formal, let us
define the matrix
$$
U = \left(
\begin{array}{cc}
0 & 1 \\ 1 & 0 
\end{array}
\right)\,,
$$ such that $U^\dag = U$ and $U^{-1} = U$. 
Then $A^\dag = U\cdot A\cdot U$, so we obtain from
Eq.~(\ref{ad-vec-eq})
\begin{equation}
A\cdot\frac{d (U\cdot{\bar y}_n)}{dr} = i\o_n^*\,U\cdot{\bar y}_n\,.  
\end{equation}
On the other hand, defining 
$$y_n = A\cdot x_n
= (r^2 (\delta \rho \; v + \rho \; \delta v),
\frac{c^2}{\r}\,\d \r + v\,{\delta v})_n,$$ 
we can write Eq.~(\ref{vec-eq}) as
\begin{equation}
A\cdot\frac{d y_n}{dr} = i\o_n\,y_n\,.  
\label{vec-y-eq}
\end{equation}
Therefore, we can identify $y_n = U\cdot{\bar y}_n$, that is, we have
an equivalence between the original equation and its adjoint.  In
consequence, we have that $\o_n = \o_n^*$.  This constitutes a proof
of the reality of eigenvalues different from those given by
Petterson, Silk \& Ostriker (1980) or Theuns
\& David (1992).

Furthermore, given the preceding identification, we can write the
orthogonality relation (\ref{orthog0}) as the orthogonality of
eigenfunctions $x_n$ or, alternately, of eigenfunctions $y_n$; we
shall use the latter, namely,
\begin{equation}
\int  y_n^* \cdot U \cdot A^{-1}\cdot y_m \,dr \propto \delta_{nm}\,.
\label{orthog}
\end{equation}
Note that the matrix $U \cdot A^{-1}$ is self-adjoint and positive 
and plays the r\^ole of a metric in the space of vectors $y$.

As we have pointed out above, the adjoint equation is valid if ${\bar
y}^*\cdot A \,x = {\bar y}^*\cdot y = {y^1}^* y^2 + {y^2}^* y^1 = 0$
at the ends of the integration interval (using the components of $y =
(y^1,y^2)$). Therefore, sufficient boundary conditions are given by
either component $y^1$ or $y^2$ vanishing at both ends. We 
choose the former to vanish, as explained in the main text.

\section{Asymptotic form of acoustic modes}
\label{appendix_B}

\subsection{Connection with WKB solution}

When the wave-length is small compared with the typical dimensions of
the zone in which waves propagate, the WKB solution is a good
approximation. The WKB solution of the differential equations for
radial modes was obtained by Petterson, Silk \& Ostriker (1980):
{\setlength\arraycolsep{2pt}
\begin{eqnarray}
\d f(r) &=& \frac{1}{\sqrt{c\,v}} \left[ A_1 \exp \left(i\o \int\! 
\frac{dr}{v+c}\right) + \right.\nonumber\\
&&\hspace{1cm}\left.
A_2 \exp \left(i\o \int\! \frac{dr}{v-c} \right) \right],\\
\d\r(r) &=& \frac{1}{r^2\sqrt{c\,v}} \left[\frac{A_1}{v+c} 
\exp\left(i\o \int\! \frac{dr}{v+c}\right) +  \right.\nonumber\\
&&\hspace{1.2cm}\left.
\frac{A_2}{v-c} \exp\left(i\o \int\! \frac{dr}{v-c} \right) \right].
\end{eqnarray}
} In this equation, the two functions that form the linear combination
with (arbitrary) coefficients $A_1$ and $A_2$ represent the outgoing
and incoming waves, respectively.  Substituting the self-similar
expressions of $v$ and $c$ and performing the integrals, we obtain the
corresponding expressions of the WKB solutions $\d f(r)$ and $\d\r(r)$. 
From them we can further obtain 
\begin{eqnarray}
\d g(r) = \frac{\s}{\a}\, r^{-1/2}  \left[ A_1 \exp 
\frac{2i\o\, r^{3/2}}{3(\b+\s)} 
- A_2 \exp 
\frac{2i\o\, r^{3/2}}{3(\b-\s)} 
\right].
\label{WKB}
\end{eqnarray}
This formula is valid for large $\o$ or, equivalently, large $r$.

The two component waves are analogous to the functions $J_{\nu}$ and
$N_{\nu}$ in the radial solution (\ref{solutionR}).  To find the
precise relation between them, it is convenient to express the radial
solution (\ref{solutionR}) in terms of the complex combinations of the
Bessel functions, namely, the Hankel functions $H^{(1)}_{\nu}$ and
$H^{(2)}_{\nu}$ (Morse \& Feshbach, 1953).  The asymptotic forms of
the Hankel functions for large values of their arguments is
\begin{eqnarray}
H^{(1)}_{\nu}(x) = \sqrt{\frac{2}{\pi x}} 
\exp\left[i\left(x-\frac{\nu\pi}{2}- \frac{\pi}{4}\right)\right] 
+ O(x^{-3/2}), \label{H1}\\
H^{(2)}_{\nu}(x) = \sqrt{\frac{2}{\pi x}} 
\exp\left[-i\left(x-\frac{\nu\pi}{2}- \frac{\pi}{4}\right)\right] 
+ O(x^{-3/2}),
\label{H2}
\end{eqnarray}
Expressing Eq.\ (\ref{solutionR}) in terms of Hankel functions (with
two new constants $C'_1$ and $C'_2$) and using the preceding
asymptotic forms, we have
\begin{eqnarray}
R(r) \sim r^{1/4} e^{-i \mu r^{3/2}} r^{-3/4}
[C'_1 \exp(i\l r^{3/2}) + C'_2 \exp(-i\l r^{3/2})],
\end{eqnarray}
still with $\mu =\frac{2 \beta \o}{3(\sigma^2-\beta^2)}$, $\l =
\frac{2 \sigma\o}{3 (\sigma^2 - \beta^2)}.$ Note that this asymptotic
form becomes $\nu$-independent and therefore {\em $l$-independent}.
Given that
$$-\mu \pm \l = 
- \frac{2 \beta \o}{3(\sigma^2-\beta^2)}  \pm 
\frac{2 \s \o}{3(\sigma^2-\beta^2)} 
= \frac{2 \o}{3(\b \pm \s)}\;,
$$
we have that 
\begin{eqnarray}
R(r) \sim r^{-1/2} \left[C'_1 \exp \frac{2 i\o\, r^{3/2}}{3(\b+ \s)}
+ C'_2 \exp\frac{2 i\o\, r^{3/2}}{3(\b - \s)}\right].
\end{eqnarray}
This last form is equivalent to the WKB form (\ref{WKB}) but 
is valid for non-radial modes as well.

\subsection{Asymptotic form of radial eigenfunctions}

\begin{figure}
\centering{\includegraphics[width=5.5cm]{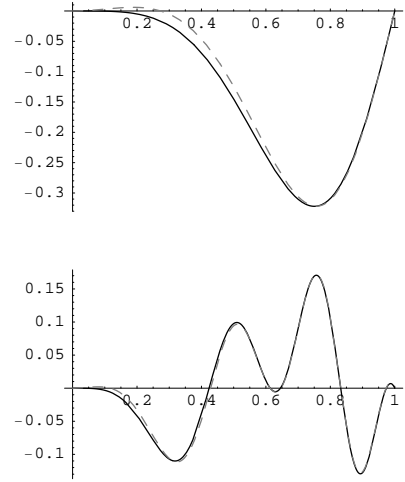}}
\caption{Real parts of radial modes with $n=1$ and $n=4$ (solid lines)
compared with their asymptotic trigonometric forms (dashed
lines). Note that the difference diminishes as $r$ grows.}
\label{asymptotic}
\end{figure}

For $n \gg 1$ we can substitute in Eqs.\ (\ref{fn}) and (\ref{gn}) the
asymptotic forms of the Bessel functions for large values of their
arguments (Morse \& Feshbach, 1953):
$$
J_\nu(x) = \sqrt{\frac{2}{\pi x}} 
\cos\left(x-\frac{\nu\pi}{2}- \frac{\pi}{4}\right)
+ O(x^{-3/2})
$$
[consistently with Eqs.\ (\ref{H1}) and (\ref{H2})].
In particular,
\begin{eqnarray*}
J_{5/6}(\l_n r^{3/2}) = \sqrt{\frac{2}{\pi \l_n}} r^{-3/4}
\cos\left(\l_n r^{3/2} -\frac{2\pi}{3}\right)
\, ,\\
J_{-1/6}(\l_n r^{3/2}) = \sqrt{\frac{2}{\pi \l_n}} r^{-3/4}
\cos\left(\l_n r^{3/2} -\frac{\pi}{6}\right)
\, .
\end{eqnarray*}
Replacing $\cos\left(\l_n r^{3/2} - 2\pi/3\right)$ with
$\sin\left(\l_n r^{3/2} - \pi/6\right)$, we can make use of the
boundary condition $\d f(1,t) = 0$ to obtain the spectrum $\l_n
\approx \left(n + 1/6\right)\pi, \; n= 1,2,\ldots$.  The corresponding
proper functions are
\begin{eqnarray}
\d f_n(r,t) \approx \exp\left[-i\pi\left(n + \frac{1}{6}\right)
\left(\M r^{3/2}+\frac{3}{2} (1
-\M^2)t\right)\right] \times \nonumber\\
r^{1/2}\, i \,\frac{\a}{\sigma}\,
\frac{1}{\pi}
\,{\sqrt{\frac{2}{n + 1/6}}}\,
\sin \left[\left(n + \frac{1}{6}\right)\pi\, r^{3/2} - \frac{\pi}{6}\right] ,
\label{fn-as}
\\ \d g_n(r,t) \approx \exp\left[-i\pi\left(n + \frac{1}{6}\right) 
\left(\M r^{3/2}+\frac{3}{2} (1
-\M^2)t\right)\right] \times \nonumber\\
r^{-1/2} \,\frac{1}{\pi}
\,{\sqrt{\frac{2}{n + 1/6}}}\,
\cos \left[\left(n + \frac{1}{6}\right)\pi\, r^{3/2} - \frac{\pi}{6}\right].
\label{gn-as}
\end{eqnarray}
In Fig.\ \ref{asymptotic} we plot two low radial modes (${\cal M} =
1/2$), that is, their real parts, to compare them with their
asymptotic expressions according to Eqs.\ (\ref{fn-as})and
(\ref{gn-as}). We note that the difference between the actual and the
asymtotic forms diminishes as $n$ or $r$ grow, but it is small even in
the most unfavourable case, namely, $n=1$ and $r < 0.5$.

The simpler form in terms of trigonometric functions allows us to
obtain an analytic form of the orthogonality of eigenfunctions:
{\setlength\arraycolsep{2pt}
\begin{eqnarray}
{\langle y_n, y_m \rangle} \approx 12 \times \hbox{\hspace{6.1cm}} 
\nonumber\\
\frac{\int\limits_0^1 e^{\frac{i }{2}\left( m - n \right) \pi
r^{\frac{3}{2}}}\, r^{\frac{1}{2}} \left( \cos (\left( m - n \right)
\pi \, r^{\frac{3}{2}}) - i \,\M \,\sin (\left( m - n \right) \pi
\,r^{\frac{3}{2}}) \right) dr }{{\pi }^2\, {\sqrt{\left( 1 + 6\,m
\right) \left(1 + 6\,n \right) }}\, \left(1 - {\M}^2 \right) }
\nonumber\\ = \delta_{nm}\, \frac{8}{\left( 1 + 6\,n \right) {\pi }^2
\left(1 - {\M}^2\right) } \,. \hbox{\hspace{3.5cm}}
\label{orthog-J-as} 
\end{eqnarray}
}
In particular, we have a useful approximation of the norm.
We have found that the norm given by this formula is in fact a good
approximation already for $n =1$, with a relative error smaller than
$0.001$ (for $\M = 1/2$), which rapidly diminishes as $n$ increases.

\end{document}